\documentclass[runningheads]{llncs}
\usepackage[T1]{fontenc}
\usepackage{graphicx,wrapfig,lipsum,booktabs}
\usepackage{cite}
\usepackage{amsmath,amssymb,amsfonts}
\usepackage{algorithmic}
\usepackage{graphicx}
\usepackage{textcomp}
\usepackage{booktabs}
\usepackage{csquotes}
\usepackage{multirow}
\usepackage{mathtools}
\usepackage{makecell}
\usepackage[normalem]{ulem}
\usepackage[pagebackref,breaklinks,colorlinks]{hyperref}
\usepackage[usenames,dvipsnames,svgnames,table]{xcolor}
\usepackage{subcaption}
\usepackage{xspace}
\usepackage{dsfont}
\makeatletter
\DeclareRobustCommand\onedot{\futurelet\@let@token\@onedot}
\def\@onedot{\ifx\@let@token.\else.\null\fi\xspace}

\def\ie{\emph{i.e}\onedot}

\makeatother
\DeclareMathOperator{\Loss}{\mathcal{L}}

\renewcommand{\vec}[1]{{\mathbf #1}}
\usepackage{pifont}
\usepackage{hyperref}
\hypersetup{
    colorlinks=true,
    linkcolor=blue,
    filecolor=magenta,      
    urlcolor=cyan,
}
\usepackage{mwe}
\usepackage[capitalize]{cleveref}
\crefname{section}{Sec.}{Secs.}
\Crefname{section}{Section}{Sections}
\Crefname{table}{Table}{Tables}

\begin{document}
\title{Bilateral Hippocampi Segmentation in Low Field MRIs Using Mutual Feature Learning via Dual-Views}
%
\titlerunning{LoFiHippSeg}
%
\author{Himashi Peiris\inst{1,2}\orcidID{0000-0003-0464-1182} \and
Zhaolin Chen\inst{1,2}\orcidID{0000-0002-0173-6090}}

%
\authorrunning{H. Peiris et al.}
%
\institute{Department of Data Science \& AI, Faculty of IT, Monash University, Melbourne, Australia. \and Monash Biomedical Imaging (MBI), Monash University, Melbourne, Australia. \\
\email{\{Himashi.Peiris, Zhaolin.Chen\}@monash.edu}}
\maketitle              
\begin{abstract}
Accurate hippocampus segmentation in brain MRI is critical for studying cognitive and memory functions and diagnosing neurodevelopmental disorders. While high-field MRIs provide detailed imaging, low-field MRIs are more accessible and cost-effective, which eliminates the need for sedation in children, though they often suffer from lower image quality. In this paper, we present a novel deep-learning approach for the automatic segmentation of bilateral hippocampi in low-field MRIs. Extending recent advancements in infant brain segmentation to underserved communities through the use of low-field MRIs ensures broader access to essential diagnostic tools, thereby supporting better healthcare outcomes for all children. 
Inspired by our previous work, Co-BioNet, the proposed model employs a dual-view structure to enable mutual feature learning via high-frequency masking, enhancing segmentation accuracy by leveraging complementary information from different perspectives. Extensive experiments demonstrate that our method provides reliable segmentation outcomes for hippocampal analysis in low-resource settings. 
The code is publicly available at: \href{https://github.com/himashi92/LoFiHippSeg}{https://github.com/himashi92/LoFiHippSeg}.

\end{abstract}
%
%
%

%
%
\section{Introdcution}
\label{sec:intro}
The hippocampus is a vital subcortical structure in memory formation and cognitive processes. Accurate hippocampus segmentation in MRI scans is essential for studying neurodevelopmental disorders and cognitive impairments. High-field MRIs, with their superior image quality, are typically used for this task, but their high cost and limited availability pose significant barriers, especially in low-resource settings~\cite{sheth2021assessment}. Low-field MRIs, while more accessible, often produce images with lower resolution and increased noise, making accurate hippocampal segmentation challenging. Recent advancements in deep learning have shown promise in improving medical image segmentation~\cite{cciccek20163d,peiris2022hybrid}. However, existing methods primarily focus on high-field MRIs, leaving a gap in effective techniques for low-field MRI segmentation. Inspired by the mutual feature learning mechanism of Co-BioNet~\cite{peiris2023uncertainty}, we propose a novel approach that utilizes a dual view structure to enhance segmentation performance in low-field MRIs. By learning complementary features from two different views by associating high-frequency images of the given low-field MR via high-frequency masking, our model can effectively capture the complex structures of the bilateral hippocampi. This approach of utilizing high-frequency images alongside the original low-field images demonstrates potential as a valuable tool for improving the usability of low-field images in low-resource settings, as it minimizes the need for extensive external tools and datasets.

This paper presents our dual-view mutual feature learning framework and demonstrates its efficacy through extensive experiments on the LISA 2024 low-field MRI dataset. Our results highlight the potential of this approach in providing accurate and reliable hippocampal segmentation, thereby facilitating better diagnostic and research capabilities in resource-constrained environments.

%
%
\section{Dataset}
\label{sec:dataset}

\begin{figure}[h!]
\scriptsize
\centering
\includegraphics[width=0.95\linewidth]{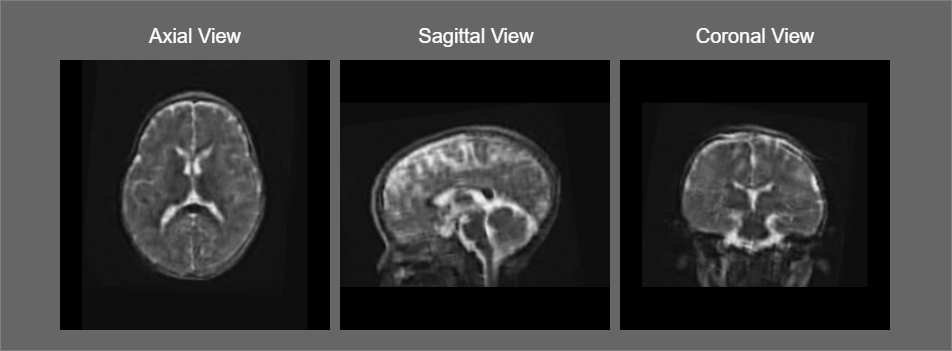}
\caption{Sample case from LISA Dataset.}
\label{fig:dataset}
\end{figure}

The dataset utilized in this study comprises high-field T2-weighted MRI scans and synchronized low-field Hyperfine scans acquired from institutions in Uganda, South Africa, and the United States~\cite{deoni2022development}. Expert MRI technicians collected the images, ensuring high-quality data. The dataset includes meticulously reviewed bilateral hippocampi segmentations by an expert medical image evaluator, providing a reliable ground truth. The images are available in NIFTI (.nii.gz) format, with low-field images registered to high-field scans through a 9-point linear registration process. Orthogonal low-field images were processed using the ANTs multivariate template construction and aligned with a pediatric T2 template, subsequently coregistered to matching high-field scans using FLIRT from the FSL toolbox~\cite{smith2004advances}. This robust dataset underpins developing and evaluating our deep-learning model for accurate hippocampal segmentation in low-field MRIs.

%
%
\section{Methodology}
\label{sec:method}

\begin{figure}[ht!]
\scriptsize
\centering
\includegraphics[width=0.95\linewidth]{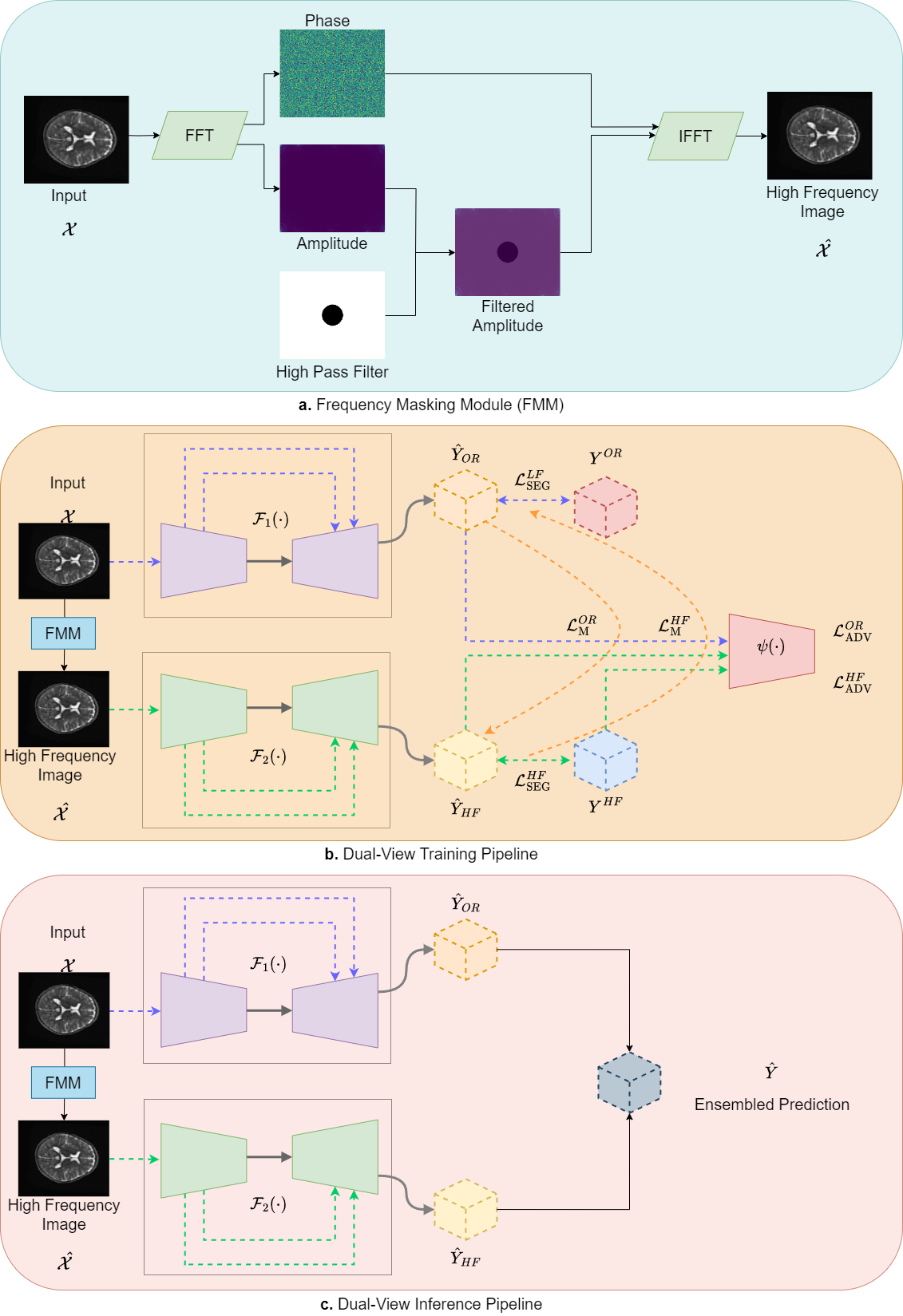}
\caption{\textbf{Overview of Proposed Dual-View Pipeline, LoFiHippSeg.} Here, $\mathcal{F}_1(\cdot)$ and $\mathcal{F}_2(\cdot)$ are structuraly similar VNet models.}
\label{fig:pipeline}
\end{figure}

\subsection{Notations \& Problem Formulation.}
In our paper, we represent vectors and matrices using bold lowercase $\mathbf{x}$ and bold uppercase $\mathbf{X}$, respectively. The norm of a vector is denoted by $\| \cdot \|$, with $\|\mathbf{x}\|_1 = \sum_i |\mathbf{x}[i]|$, where $\mathbf{x}[i]$ signifies the element at position $i$ in $\mathbf{x}$. The inner product between vectors is denoted by $\langle \cdot,\cdot \rangle$, and $\|\mathbf{x}\|_2^2 = \langle \mathbf{x},\mathbf{x} \rangle$. When norms and inner products are applied to 3D tensors, we assume the tensors are flattened. For instance, for 3D tensors $\mathbf{A}$ and $\mathbf{B}$, $\langle \mathbf{A},\mathbf{B} \rangle = \sum_{i,j,k} \mathbf{A}[i,j,k]\mathbf{B}[i,j,k]$ and $\| \mathbf{A} \|_1 = \sum_{i,j,k} |\mathbf{A}[i,j,k]|$.

Consider a dataset $\mathcal{X}_1 = \{(\mathbf{X}_i,\mathbf{Y}_i)\}_{i=1}^n$ consisting of $n$ samples, where each sample $(\mathbf{X}_i,\mathbf{Y}_i)$ includes an image $\mathbf{X}_i \in \mathbb{R}^{C \times H \times W \times D}$ and its corresponding ground-truth segmentation mask $\mathbf{Y}_i \in \{0,1\}^{K \times H \times W \times D}$, encoded as a one-hot $K$-dimensional vector for a $K$-class problem per voxel. Here, $C$, $H$, $W$, and $D$ denote the number of channels, height, width, and depth of the input medical volume. Similarly, in order to create Dual-Views of the input, we use a high pass filtering method to generate a high frequency of the low-field MR volume, which creates another dataset $\mathcal{X}_2 = \{(\hat{\mathbf{X}}_i,\mathbf{Y}_i)\}_{i=1}^n$ consisting of $n$ samples. The primary objective is to co-learn segmentation models from $\mathcal{D} = \mathcal{X}_1~\cup~\mathcal{X}_2$.

\subsection{LoFiHippSeg Architecture.}
Inspired by our previous works~\cite{peiris2021duo,peiris2023uncertainty}, we propose a Dual-View deep learning architecture named \textbf{LoFiHippSeg} to learn features from \textbf{Lo}w-\textbf{Fi}eld MRIs for \textbf{Hipp}ocampi \textbf{Seg}mentaion. As shown in the conceptual diagram in~\cref{fig:pipeline}, we use two segmentation networks denoted as $\mathcal{F}_1(\cdot)$ and $\mathcal{F}_2(\cdot)$, which creates Dual-Views which mutually learn from complementary features. $\mathcal{X}_1$ dataset and $\mathcal{X}_2$ is used to train two segmentation models, respectively.
Considering the computational complexity in our pipeline, we use VNet~\cite{vnet} as the segmentation model for dual-view training. As the critic network, we use a fully convolutional neural network similar to encoder architecture, following recent works~\cite{peiris2023uncertainty}.

\subsubsection{Frequency Masking Module (FMM).}
Frequency domain analysis is crucial in medical imaging, including MRI reconstruction and image-denoising applications. The low frequencies in an image’s Fourier spectrum represent the mean image intensity (DC signal) and the intensities of significant image components. Conversely, high frequencies capture fine details such as edges, boundaries between tissues, and the delicate outlines of structures~\cite{liang2024itermask2}. Considering this, we used the data augmentation based on the frequency masking approach to create another view of the original low-field MR volume. We employ both the original low-field MR volume and its high-frequency components to train our dual views based on the principle of consensus~\cite{xu2013survey,dasgupta2002pac,blum1998combining}. This approach ensures that the complementary information from both views is integrated, enhancing the overall performance of the model. 
The original low-field MR volume provides essential structural and intensity information, capturing the general anatomy and major tissue contrasts. In contrast, the high-frequency components emphasize fine details such as edges and boundaries, which are crucial for accurately delineating structures. By training with both views, the network can leverage the strengths of each, leading to more robust and precise segmentation. The consensus between these dual views helps reinforce consistent and accurate predictions, ultimately improving the model’s reliability and effectiveness in various medical imaging tasks.

Consider a low-field MR volume $\mathbf{X} \in \mathbb{R}^{C \times H \times W \times D}$. To perform frequency masking, we first transform $\mathbf{X}$ to the frequency domain using the Fourier transform $\mathcal{FFT}$.
\begin{equation}
    \mathbf{X}_f = \mathcal{FFT}(\mathbf{X})
\end{equation}

The Fourier-transformed medical volume $\mathbf{X}_f$ can be decomposed into its amplitude $\mathbf{A}_{\mathbf{X}_f}$ and phase $\mathbf{P}_{\mathbf{X}_f}$ components. Next, we define a high-pass filter $\mathbf{H} \in \{0, 1\}^{D \times H \times W}$ that selectively retains high-frequency components. The mask is designed such that it zeros out the low-frequency components and retains the high-frequency ones:
\begin{equation}
    \mathbf{A}_{\mathbf{X}_f}^{\text{high}} = \mathbf{A}_{{\mathbf{X}_f}} \odot \mathbf{H}
\end{equation}

Here, $\odot$ denotes the element-wise multiplication. The phase remains unchanged: The high-frequency representation $\hat{\mathbf{X}}_{\text{high}}$ in the frequency domain is then:
\begin{equation}
    \hat{\mathbf{X}}_{\text{high}} = \mathbf{A}_{\mathbf{X}_f}^{\text{high}} \cdot e^{i \mathbf{P}_{\mathbf{X}_f}}
\end{equation}

This high-frequency representation is then transformed back to the spatial domain using the inverse Fourier transform $\mathcal{FFT}^{-1}$:
\begin{equation}
    \hat{\mathbf{X}} = \mathcal{FFT}^{-1}(\hat{\mathbf{X}}_{\text{high}})
\end{equation}

The resulting $\hat{\mathbf{X}}$ captures the high-frequency components of the original image volume $\mathbf{X}$.

\subsection{Objective Function.}
Building on our previous works~\cite{peiris2021duo,peiris2023uncertainty}, we train each segmentation network (VNet model), by optimizing the following min-max problem:
\begin{align}
    \label{eqn:min_max_overall_loss}  
    \min_{\theta_i}\max_{\theta_c} ~\Loss^i(\vec{\Theta}; \mathcal{D}) \;.
\end{align}
Here, $\vec{\Theta}$ includes all the networks' parameters, \ie, ${\theta}_1,{\theta}_2,\theta_c$. The min-max problem in Equation~\cref{eqn:min_max_overall_loss} aims to determine whether the prediction masks generated by the segmentation networks belong to the same distribution as the ground truth or if they deviate from it. Here, both original low-field MRI and High-frequency Images of low-field MRI medical volumes ($\mathcal{D}$) are utilized simultaneously during training using the following multi-task loss function:
\begin{align}
    \label{eqn:loss1}
    \Loss^i({\theta}_i; \mathcal{X}_i) \coloneqq 
    \Loss^i_{\mathrm{SEG}}({\theta}_i; \mathcal{X}_i) 
    + \lambda_m \Loss^i_{\mathrm{M}}({\theta}_i; \mathcal{X}_i) 
    + \lambda_c \Loss^i_{\mathrm{ADV}}({\theta}_i; \mathcal{X}_i)\;, 
\end{align}
where $\Loss^i_{\mathrm{SEG}}$, $\Loss^i_{\mathrm{M}}$ and $\Loss^i_{\mathrm{ADV}}$ denote the Segmentation loss, the Masked Spatial Cross-Entropy loss, and the Adversarial loss, respectively.
Here, $\lambda_m$ and $\lambda_c$ denote weighted parameters to control individual loss terms during the training.  We set $\lambda_m=0.3$ and $\lambda_c=0.01$ in all our experiments.

\subsubsection{The Segmentation Loss ($\Loss_{\mathrm{SEG}}$).} 
The Segmentation loss drives each segmentation network to produce prediction masks for labeled data that closely match the ground truth masks. We define the total Segmentation loss as the sum of the Cross-Entropy loss and Dice loss, both calculated voxel-wise. The primary segmentation loss is defined as follows:
\begin{align}
    \label{eqn:ce_loss}
    \Loss^i_{\text{CE}}(\theta_i;\mathcal{X}_i) &= \sum_{i \in m}\Bigg[ -\mathbb{E}_{(\vec{X}, \vec{Y}) \sim \mathcal{X}_i} \Big[ 
     \big\langle \vec{Y}, \log \big(\mathcal{F}_i(\vec{X}, i, \hat{\vec{Y}}_{i-1}, \vec{Z}_{i-1}) \big) \big\rangle \Big] \Bigg],\\
     \label{eqn:dice_loss}
    \Loss^i_{\text{DICE}}(\theta_i;\mathcal{X}_i) &= \sum_{i \in m} \Bigg[ 1 - \mathbb{E}_{(\vec{X}, \vec{Y}) \sim \mathcal{X}_i} \Bigg[ \frac{ 2 \big \langle \vec{Y}~,~\mathcal{F}_i(\vec{X}, i, \hat{\vec{Y}}_{i-1}, \vec{Z}_{i-1}) 
      \big \rangle}
     {\big\|\vec{Y}\big\|_1 + \big\|\mathcal{F}_i(\vec{X}, i, \hat{\vec{Y}}_{i-1}, \vec{Z}_{i-1})\big\|_1} \Bigg]\Bigg],
\end{align}
\begin{align}
\label{eqn:sup_loss}
     \Loss^i_{\mathrm{SEG}}({\theta}_i; \mathcal{X}_i) &= \Loss^i_{\text{CE}}({\theta}_i;\mathcal{X}_i) + \Loss^i_{\text{DICE}}(\theta_i;\mathcal{X}_i),
\end{align}
\vspace{-8mm}
\subsubsection{The Adversarial Loss ($\Loss_{\mathrm{ADV}}$).} 
In our training pipeline, we use a critic network which has the functionality of $\psi:[0,1]^{H \times W \times D} \to [0,1]^{H \times W \times D}$ that helps the segmentation network to generate realistic segmentation masks using min-max game as defined in~\cref{eqn:min_max_overall_loss}. The adversarial loss for the training segmentation network is defined as:
\begin{align}
    \Loss^i_{\mathrm{ADV}}(\theta_i; \mathcal{X}) &\coloneqq -\mathbb{E}_{(\vec{X},\vec{Y} \sim \mathcal{X}_i)} \Big[\sum_{a \in H} \sum_{b \in W}\sum_{c \in D}  
    \log\big(\psi(\mathcal{F}_i(\vec{X}, \hat{\vec{Y}}))[a,b,c]\big) 
    \Big]\;,
    \label{eqn:loss_gen_adv}
\end{align}

\subsubsection{The Masked Spatial Loss ($\Loss_{\mathrm{M}}$).} 
Further, we integrate a spatial masked CE loss to train the model via uncertainty, which leads to co-learn from each model's features. Here, we make the masked segmentation prediction map by binarizing the confidence map using a predefined threshold of $T=0.2$. The masked loss is defined as follows:
\begin{align}
\scriptsize
    \Loss^i_{\mathrm{M}}(\theta_i; \mathcal{X}_i) &\coloneqq -\mathbb{E}_{(\vec{X},\vec{Y} \sim \mathcal{X}_i)} \Big[ \sum_{\substack{a , b , c }} 
    \mathbf{1}\big( \psi(\mathcal{F}(\vec{X}, \hat{\vec{Y}})[a,b,c] > \text{T} \big) \nonumber\\ &
    \vec{Y}[a,b,c] 
    \log \big(\mathcal{F}(\vec{X}, \hat{\vec{Y}})[a,b,c] \big) 
   \Big]\;,
    \label{eqn:loss_mask}
\end{align}

\subsubsection{The Critic Loss ($\Loss_{\mathrm{C}}$).} 
To train the critic network, we use segmentation masks and their ground truth masks. We define the adversarial loss as maximizing the log-likelihood as:
\begin{align}
    \Loss_{\mathrm{C}}(\theta_c; \mathcal{D}) \coloneqq \mathbb{E}_{(\vec{X}, \vec{Y}) \sim \mathcal{X}_1} \bigg[&\sum_{a \in H} \sum_{b \in W} \sum_{c \in D} \Big\{\eta \log\big(  \psi(\vec{Y})[a,b,c]\big)+ (1-\eta)
    \notag \\&
     \log\big(1 - \psi_i(\mathcal{F}_i(\vec{X}, \hat{\vec{Y}}))[a,b,c]\big) 
    \Big\}\bigg] + \notag \\&
    \mathbb{E}_{(\vec{X}, \vec{Y}) \sim \mathcal{X}_2} \bigg[\sum_{a \in H} \sum_{b \in W} \sum_{c \in D} \Big\{\eta \log\big(  \psi(\vec{Y})[a,b,c]\big)
    \notag \\&
    + (1-\eta)\log\big(1 - \psi_i(\mathcal{F}_i(\vec{X}, \hat{\vec{Y}}))[a,b,c]\big) 
    \Big\}\bigg]\;.
    \label{eqn:loss_critic}
\end{align}
where $\eta = 0$ when the sample is a prediction mask from a segmentation network, and $\eta = 1$ when the sample is obtained from the ground truth label distribution.

%
%
\section{Experiments}
\label{sec:experiments}

\subsection{Implementation Details}
The min-max scaling was performed to standardize all volumes, followed by clipping intensity values. Images were then cropped to a fixed patch size of $128 \times 128 \times 128$ by removing unnecessary background pixels~\cite{peiris2021volumetric,peiris2022reciprocal,peiris2022hybrid}. 
The LoFiHippSeg model is implemented in PyTorch and trained using a single NVIDIA A100 GPU with 80GB of memory. For training the segmentation networks, we utilized the batch size of 4 and the SGD optimizer with a learning rate of 0.01 and a momentum of 0.9. The critic network was trained with the AdamW optimizer, which had a learning rate of 0.0001. We applied a cosine annealing scheduler to all networks throughout the training process. The training was conducted alternately between the segmentation networks and the critic. The critic is not used during inference, thereby avoiding additional computational overhead.
We split the training dataset into a training set (76\%) of 60 MR volumes for training and a validation set (24\%) of 19 MR volumes for validation. The best-performing model for the validation set is saved as the best model for official validation and testing phase evaluation.
The LISA 2024 validation dataset contains 12 MR volumes, and the Synapse portal conducts the evaluation. In the inference phase, the original volume was re-scaled using min-max normalization scaling and fed forward through the LoFiHippSeg model. The LoFiHippSeg model uses ensembled prediction from dual views as the final prediction during inference.

\subsection{Evaluation Metrics}
We will utilize five metrics for evaluating hippocampi segmentation predictions: Dice Similarity Coefficient (DSC), Hausdorff Distance (HD), HD95, Average Symmetric Surface Distance (ASSD), and Relative Volume Error (RVE). These metrics will be computed separately for the left and right hippocampus, and the results will be averaged for each patient case~\cite{maier2020bias}.

\begin{table}[]
\centering
\caption{Validation Phase Quantitative Comparison with VNet.}
\resizebox{0.8\textwidth}{!}{
\begin{tabular}{|l|c|c|c|c|c|c|}
\hline
\multirow{2}{*}{~Metric~} & \multicolumn{3}{c|}{VNet~\cite{vnet}} & \multicolumn{3}{c|}{LoFiHippSeg} \\ \cline{2-7}
& Left & Right & Average & Left & Right & Average \\
\hline
DSC & 0.69±0.23 & 0.72±0.15 & 0.70±0.19 & \textbf{0.70±0.23} &	\textbf{0.74±0.15} &	\textbf{0.72±0.19} \\

HD & ~10.56±16.78~ & 3.65±1.02 & 7.10±8.32 & \textbf{5.99±9.12} &	\textbf{3.49±1.18} &	\textbf{4.74±4.62} \\

HD95 & \textbf{2.11±1.95} & 1.88±0.78 & \textbf{1.99±1.33} & 2.16±2.03 &	\textbf{1.86±0.92}	 & 2.01±1.45 \\

ASSD & 0.87±1.19 & ~~0.61±0.39~~ & ~~0.74±0.78~~ & ~~\textbf{0.87±1.33}~~ &	~~\textbf{0.59±0.44}~~	& ~~\textbf{0.73±0.88}~~
 \\

RVE & 0.18±0.10	& \textbf{0.14±0.10} & 0.16±0.07 & \textbf{0.14±0.12}	& 0.14±0.12	 & \textbf{0.14±0.10} \\
\hline
\end{tabular}
}
\label{tab:quantitative}
\end{table}

\begin{table}[h!]
    \centering
    \resizebox{0.7\textwidth}{!}{
    \begin{tabular}{ccc}
        \textbf{Input} & \textbf{VNet~\cite{vnet}} & \textbf{LoFiHippSeg} \\
        
        \includegraphics[width=0.22\textwidth]{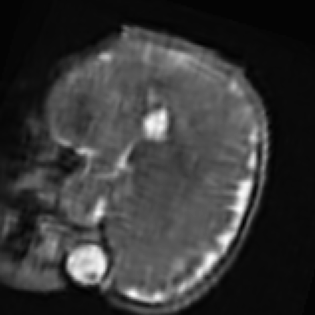} &
        \includegraphics[width=0.22\textwidth]{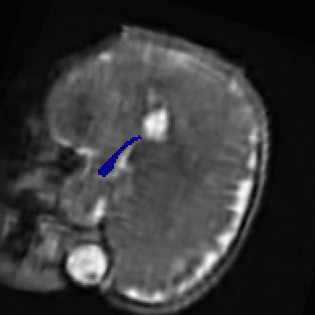} &
        \includegraphics[width=0.22\textwidth]{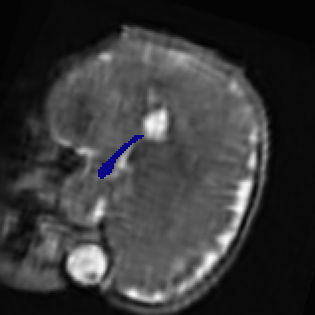} \\
        
        \includegraphics[width=0.22\textwidth]{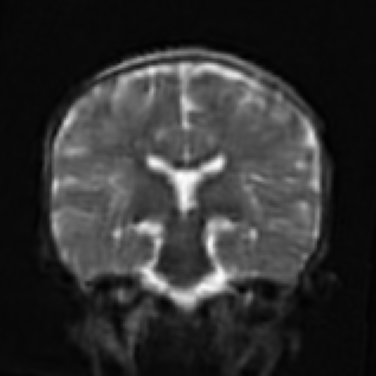} &
        \includegraphics[width=0.22\textwidth]{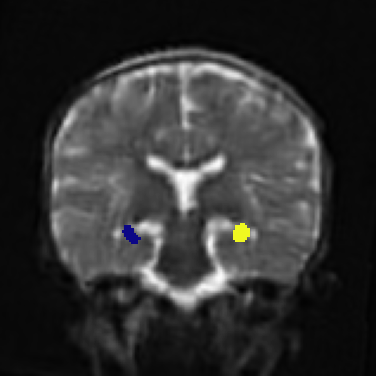} &
        \includegraphics[width=0.22\textwidth]{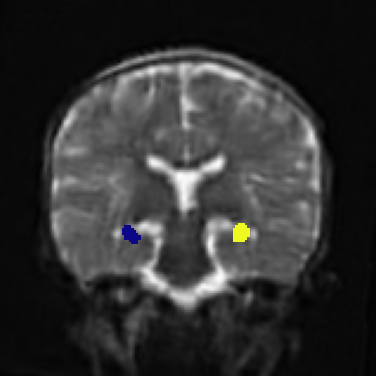} \\
        
        \includegraphics[width=0.22\textwidth]{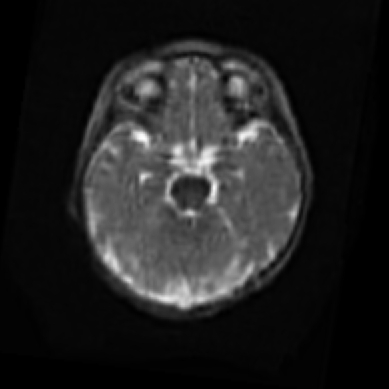} &
        \includegraphics[width=0.22\textwidth]{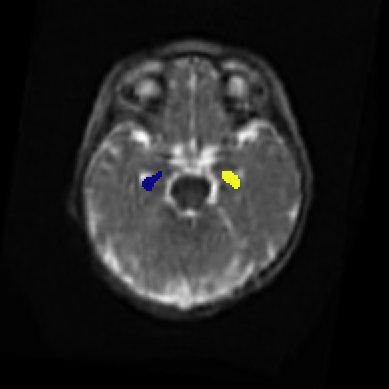} &
        \includegraphics[width=0.22\textwidth]{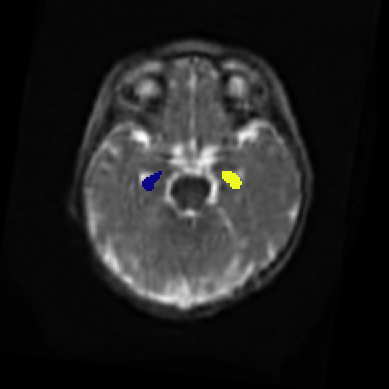} \\ 
    \end{tabular}}
    \caption{Comparison of segmentations by VNet and LoFiHippSeg across axial, sagittal, and coronal views during Validation phase.}
    \label{fig:segmentation_comparison} 
\end{table}

\begin{table}[h!]
\centering
\caption{Validation Phase Quantitative Analysis on Dual-Views of LoFiHippSeg.}
\resizebox{0.8\textwidth}{!}{
\begin{tabular}{|l|c|c|c|c|c|c|}
\hline
\multirow{2}{*}{~Metric~} & \multicolumn{3}{c|}{LoFiHippSeg View 1} & \multicolumn{3}{c|}{LoFiHippSeg View 2} \\ \cline{2-7}
& Left & Right & Average & Left & Right & Average \\
\hline
DSC & 0.69±0.23 & 0.73±0.16 & 0.71±0.19 & 0.69±0.24 & 0.73±0.13	 & 0.71±0.18 \\

HD & 8.21±15.89	& 3.52±0.98 &	5.86±7.92 & 5.80±8.03 &	6.61±11.06	& 6.20±6.43 \\

HD95 & 2.21±2.20 &	1.88±0.89 &	2.04±1.52 & 2.29±2.02 &	1.92±1.04& 2.10±1.51  \\

ASSD &  ~~0.94±1.43~~ &	~~0.59±0.45~~ & ~~0.77±0.94~~ & ~~0.89±1.28~~ &	~~0.63±0.40~~ & ~~0.76±0.83~~
\\

RVE &  0.19±0.12 &	0.15±0.12 &	0.17±0.08 & 0.15±0.09 &	0.12±0.11 &	0.14±0.08  \\
\hline
\end{tabular}
}
\label{tab:quantitative_duals}
\end{table}

\subsection{Experimental Results}
We evaluated the method's performance using the LISA 2024 Validation Phase evaluation portal, and results are shown in~\cref{tab:quantitative}. From the results, it can be seen that the proposed dual-view setting helps in better feature retrieval over a single VNet segmentation model (See~\cref{tab:quantitative_duals}). Qualitative comparison of generated prediction masks are illustrated in the~\cref{fig:segmentation_comparison}.

\subsubsection{Comparison with VNet.}
The performance comparison between LoFiHippSeg and VNet~\cite{vnet}, shown in~\cref{tab:quantitative}, indicates a modest improvement in segmentation accuracy for the LoFiHippSeg model. Specifically, LoFiHippSeg achieved a slightly higher DSC (0.72±0.19) compared to VNet (0.70±0.19), demonstrating an enhanced ability to correctly classify hippocampal regions, particularly for the right hippocampus (0.74±0.15 versus 0.72±0.15 in VNet). This increase, though marginal, signifies that the LoFiHippSeg model can better delineate hippocampal structures, possibly due to its enhanced learning capabilities from low-field MRI scans. 
In terms of boundary accuracy, LoFiHippSeg also outperformed VNet in HD (4.74±4.62 versus 7.10±8.32 for VNet). The reduction in HD suggests that the segmentation boundaries produced by LoFiHippSeg are more precise, particularly for the left hippocampus, where the HD decreased from 10.56±16.78 in VNet to 5.99±9.12 in LoFiHippSeg. This reduction could imply fewer outliers in the boundary predictions by the LoFiHippSeg model. However, HD95 showed relatively comparable values between the two models, indicating that extreme outliers in the segmentation were not substantially different. 
Regarding ASSD, which measures the average surface distance between the predicted and true segmentations, both models performed similarly, with the overall averages almost identical (0.73±0.88 for LoFiHippSeg and 0.74±0.78 for VNet). This metric aligns with the HD observations, indicating that while the general boundary accuracy has improved, there is room for further refinement. 
The RVE, a volumetric measure, shows slight improvement for LoFiHippSeg (0.14±0.10) compared to VNet (0.16±0.07). This indicates that LoFiHippSeg produces more accurate volume estimations, which is critical for clinical applications where hippocampal volume is a biomarker for neurodegenerative conditions.
As illustrated in~\cref{fig:segmentation_comparison}, the qualitative differences in segmentations are not immediately noticeable to the human eye. However, it is evident that LoFiHippSeg accurately captured some misclassified regions compared to the single VNet.

\subsubsection{Dual-View Architecture Analysis.}
\cref{tab:quantitative_duals} provides further insights into the performance of the dual-view architecture of LoFiHippSeg. Both views exhibit similar performance across most metrics, indicating robustness in the model's segmentation ability regardless of the view utilized. DSC values for both views are nearly identical (0.71±0.19 for View 1 and 0.71±0.18 for View 2), which highlights the stability of the model's segmentation performance from different perspectives. 
One key observation is the difference in HD between the two views. View 1 exhibits a lower HD (5.86±7.92) compared to View 2 (6.20±6.43). While this difference is not substantial, it suggests that the first view may provide slightly more precise boundary delineation, particularly for the left hippocampus, which shows a notable decrease in HD for View 1 (8.21±15.89 versus 5.80±8.03 for View 2). The HD95 and ASSD metrics, however, remain consistent across both views, reinforcing the robustness of the segmentation performance. 
Interestingly, the RVE metric shows a marginal improvement in View 2 (0.14±0.08) compared to View 1 (0.17±0.08). This could indicate that the second view is more effective in achieving accurate volumetric estimations, particularly for the right hippocampus (0.12±0.11 for View 2 versus 0.15±0.12 for View 1). These complementary strengths of each view suggest that a combined approach leveraging both views could potentially yield even better performance.

\subsection{Ablation Study}
\begin{figure}[h!]
\scriptsize
\centering
\includegraphics[width=0.9\linewidth]{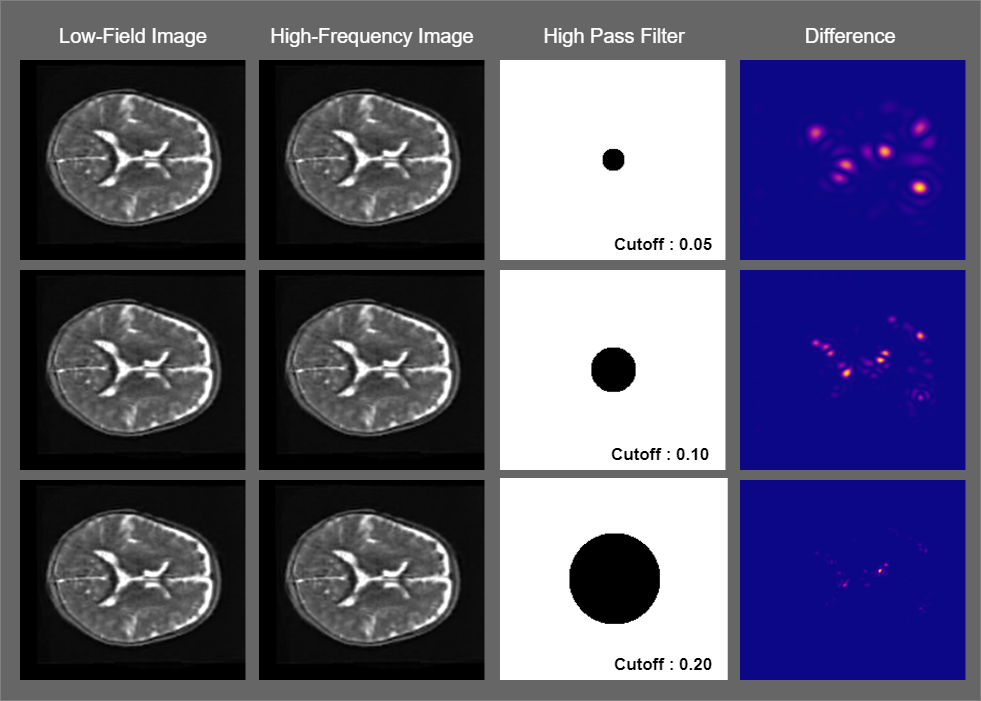}
\caption{Feature difference between Low-field Image and High-frequency Image.}
\label{fig:cutoff}
\end{figure}

One of the central claims of the proposed model is its use of high-frequency masking to generate high-frequency images of low-field images. In~\cref{fig:cutoff}, we illustrate how varying cutoff values produce different high-pass filters and the corresponding feature difference maps between the low-field and high-frequency images. These qualitative visualizations show that the feature difference decreases as the cutoff value increases. A more noticeable feature difference emerges at lower cutoff values, even though it may not be easily perceived by the naked eye, as demonstrated with a cutoff of 0.05. However, fine details are not as clearly visible at this value as they are with a cutoff of 0.1. In our ablation study, we evaluated the model's performance using these three cutoff values, with the results summarized in~\cref{tab:cutoff}. The findings indicate that a cutoff value of 0.1 yields the best performance compared to the other two.

\begin{table}[h!]
\centering
\caption{Ablation Study of cutoff value.}
\resizebox{0.6\textwidth}{!}{
\begin{tabular}{|l|c|c|c|}
\hline
\makecell{Metric} & ~~Cutoff=0.05~~ & ~~Cutoff=0.10~~ & ~~Cutoff=0.20~~ \\ 
\hline

Average DSC & 0.71±0.18  & 0.72±0.19  & 0.70±0.19   \\

Average HD & 9.12±12.58  & 4.74±4.62  & 5.90±7.93  \\

Average HD95 & 2.04±1.54  & 2.01±1.45  &  2.16±1.59 \\

Average ASSD & 0.75±0.85  & 0.73±0.88  & 0.77±0.91 \\

Average RVE & 0.14±0.10  & 0.14±0.10  &  0.20±0.10   \\

\hline
\end{tabular}
}
\label{tab:cutoff}
\end{table}

\subsection{Discussion}
The proposed LoFiHippSeg outperforms a single VNet trained on low-field MRI images. While the model achieves better results, it does present certain limitations, such as increased computational complexity. However, with the ongoing technological advancements, computational complexity is becoming less of a constraint. We believe that incorporating more advanced segmentation models over VNet could further enhance segmentation performance.

%
%
\section{Conclusion}
\label{sec:conclusion}
In this study, we introduced a novel deep-learning approach for the automatic segmentation of bilateral hippocampi in low-field MRIs, addressing a critical need in diagnosing and studying cognitive and memory functions in neurodevelopmental disorders. By adapting recent advancements in infant brain segmentation to low-field MRIs, our method extends the accessibility of essential diagnostic tools to underserved communities, promoting equitable healthcare for all children.

\begin{credits}
\subsubsection{\ackname} This study was funded by the Australian Research Council Discovery Program DP210101863.

\subsubsection{\discintname}
The authors have no competing interests. 

\end{credits}
%
%

\bibliographystyle{splncs04}
\bibliography{refernces}
\end{document}